**Title:** The Sharpest Ultraviolet view of the star formation in an extreme environment of the nearest Jellyfish Galaxy IC 3418

**Authors:** Ananda Hota[1,*], Ashish Devaraj[2], Ananta C. Pradhan[3], C S Stalin[2], Koshy George[4], Abhisek Mohapatra[3], Soo-Chang Rey[5], Youichi Ohyama[6], Sravani Vaddi[7], Renuka Pechetti[8], Ramya Sethuram[2], Jessy Jose[9], Jayashree Roy[10], Chiranjib Konar[11]

1. #eAstroLab , UM-DAE Centre for Excellence in Basic Sciences, University of Mumbai, Mumbai, India
2. Indian Institute of Astrophysics, Koramangala II Block, Bangalore, India
3. Department of Physics and Astronomy, National Institute of Technology, Rourkela, Odisha 769008, India
4. Faculty of Physics, Ludwig-Maximilians-Universität, Scheinerstr 1, Munich, 81679, Germany
5. Chungnam National University, Daejeon 34134, Republic of Korea
6. Academia Sinica Institute of Astronomy and Astrophysics, PO Box 23-141, Taipei 106, Taiwan
7. Arecibo Observatory, NAIC, HC3 Box 53995, Arecibo, Puerto Rico, PR 00612, USA
8. Astrophysics Research Institute, Liverpool John Moores University, 146 Brownlow Hill, Liverpool L3 5RF, UK
9. Indian Institute of Science Education and Research (IISER) Tirupati, Tirupati 517507, India
10. Inter-University Centre for Astronomy and Astrophysics (IUCAA), Pune, 411007, India
11. Amity Institute of Applied Sciences, Amity University Uttar Pradesh, Sector-125, Noida-201303, India

 * Corresponding author. E-mail: hotaananda@gmail.com

**Abstract:** We present the far ultraviolet (FUV) imaging of the nearest Jellyfish or Fireball galaxy IC3418/VCC 1217, in the Virgo cluster of galaxies, using Ultraviolet Imaging Telescope (UVIT) onboard the ASTROSAT satellite. The young star formation observed here in the 17 kpc long turbulent wake of IC3418, due to ram pressure stripping of cold gas surrounded by hot intra-cluster medium, is a unique laboratory that is unavailable in the Milkyway. We have tried to resolve star forming clumps, seen compact to GALEX UV images, using better resolution available with the UVIT and incorporated UV-optical images from Hubble Space Telescope archive. For the first time, we resolve the compact star forming clumps (fireballs) into sub-clumps and subsequently into a possibly dozen isolated stars. We speculate that many of them could be blue supergiant stars which are cousins of SDSS J122952.66+112227.8, the farthest star (~17 Mpc) we had found earlier surrounding one of these compact clumps. We found evidence of star formation rate (4 - 7.4 x $10^{-4}$ M☉ $yr^{-1}$) in these fireballs, estimated from UVIT flux densities, to be increasing with the distance from the parent galaxy. We propose a new dynamical model in which the stripped gas may be developing vortex street where the vortices grow to compact star forming clumps due to self-gravity. Gravity winning over turbulent force with time or length along the trail can explain the puzzling trend of higher star formation rate and bluer/younger stars observed in fireballs farther away from the parent galaxy.

**1.1 Introduction:** The basic understanding of the star formation process has still been a puzzle (McKee & Ostriker 2007, Zinnecker & Yorke, 2007, Kennicutt & Evans 2012, Krumholz 2014 ). The high resolution optical observations with the Hubble Space Telescope (HST), Ultraviolet (UV) observations with the GALEX, Infrared observations with the Spitzer and recently cold molecular gas and dust observations with the Atacama Large Millimeter/submillimeter Array (ALMA) have been providing crucial missing pieces towards resolving this age-old puzzle. Further this has been assisted with modern magneto-hydrodynamic numerical simulation studies to achieve a correct theoretical understanding of star formation (e.g. Wu et al. 2017). The nearest dwarf galaxies, Small Magellanic Cloud (SMC) and Large Magellanic Cloud (LMC), having little dust and gas or almost transparent, have been providing unique opportunities to investigate star formation at high resolution (Oey & Massey (1995), Indu & Subramaniam (2011)). Triggered star formation has been commonly observed in our Milkyway galaxy and, LMC and SMC where gaseous flow of the interstellar medium (spiral arm density wave or supernova shell expansion) is triggering further young star formation (Elmegreen & Elmegreen (1986), Oey & Massey (1995)). Galaxy merger has also been a well-documented triggering mechanism as observed in interacting, merging and Ultra-Luminous Infra-Red Galaxies (ULIRGs) with massive central starbursts (Jog & Solomon 1992, Sanders & Mirabel 1996). Role of various instabilities, turbulence (both radiation and hydrodynamically driven) and magnetic field are certainly critical to star formation though it complicates the matter significantly (Federrath & Klessen 2012, Wu et al. 2017, Müller et al. 2020).

We bring attention to triggered star formation in external galaxies in two types of extreme but opposite environments which are not only rare but also can not be observed in the Milkyway or LMC/SMC. They are so unique opportunities that it is almost like our understanding of star formation is put to extreme tests. First case is when the relativistic radio jet of non-thermal plasma, emitting synchrotron radiation, from the accretion on to the supermassive black hole, hits the outer cold gas (seen in H$\textsc{i}$ and CO emission lines) in a galaxy and triggers young star formation as seen in Hα line or blue optical light and UV emission. This case, also referred to as 'positive feedback', is seen in the nearest radio galaxy Cen A, Death-star galaxy, Minkowski Object etc. (Salome et al. 2016, Santoro et al. 2015, Croft et al. 2006, Evans et al. 2008, Zavaro et al. 2020). This is in contrast to the well-known negative feedback that is required for AGN-feedback models to explain black hole galaxy co-evolution (e.g. Springel, Di Matteo & Hernquist 2005, Croton et al. 2006, Hopkins 2006, Fabian 2012, Hota et al. 2012, 2016, Hardcastle & Croston 2020). The second unique case, focus of this manuscript, is in the 'Jellyfish' or 'Fireball' galaxy where the triggering mechanism is almost opposite. Here the cold Interstellar medium (ISM) of the galaxy has been ram pressure stripped by the million degree hot, X-ray emitting, thermal plasma of the Intra-cluster Medium (ICM) through which the galaxy has been moving with nearly a thousand km s$^{-1}$ velocity (Gunn & Gott 1972, Abadi, Moore & Bower 1999, Oosterloo & van Gorkom 2005, Vollmer et al. 2001). In some rare cases, the ongoing star formation in the striped gas tail of these systems along with the disk of the galaxy gives the appearance of a Jellyfish in the sky. Here the cold gas and the surrounding thermal plasma are similarly in high relative velocity and in extreme contrast of physical composition and state of matter. The number of jet-triggered star formation can be observed is nearly six radio galaxies and cases of Jellyfish galaxies could be nearly a dozen (GASP survey; Poggianti et al. 2017). We had proposed to observe two such targets NGC3801 with radio-lobe feedback (Hota et al. 2012) and IC3418 the nearest Jellyfish galaxy with the ASTROSAT for best possible angular and spatial resolution. IC3418 was observed with UVIT onboard the ASTROSAT satellite and we report our study briefly in this manuscript focusing on the compact star forming clump or fireballs.

**1.2 The Jellyfish or Fireball galaxy IC3418:**
Numerous examples of ram pressure-stripped gas tails with H$\textsc{i}$ and Hα emission, extending up to 100 kpc, have been discovered. However, only recently they have been confirmed to continue forming young stars in clumps of these tails. As the clumps become bright in UV, blue optical light and Hα, they create stunning "Jellyfish" galaxy structures with clumps having "Fireball" like impression where the small tails of the fireballs are seen oppositely directed to the large tail of the galaxy (Cortese et al. 2007, Sun et al. 2007, Yoshida et al. 2008, Smith et al. 2010, Hester et al. 2010, Yagi et al. 2013, Poggianti et al. 2017, Gullieuszik et al. 2017, George et al. 2018). As seen in Fig. 1 (top right panel), IC3418 has a clumpy star forming 17 kpc long tail which is bright in GALEX UV images (Hester et al. 2010). IC3418 being the closest (1" ~ 80 pc for 16.7 Mpc to Virgo cluster (Kenney et al. 2014)), it has enabled a detailed star formation study of its tail (Hester et al. 2010, Fumagalli et al. 2011, Kenney et al. 2014). We had independently discovered this in 2007 and have been following up with Lulin optical imaging, Subaru multi-slit spectroscopy and stellar population synthesis analyses. We had reported the discovery of the farthest single star SDSS J122952.66+112227.8, a blue supergiant star in its tail using Subaru spectroscopy and CFHT imaging (Ohyama & Hota 2013). Several puzzles of the star forming clumps in the tail or "fireballs" are yet to be resolved. UV and optical colour of the fireballs farther away from the parent galaxy are found to be relatively bluer than the nearby fireballs as well as the parent galaxy (Hester et al. 2010, Yoshida et al. 2008, Fumagalli et al. 2011, Kenney et al 2014). From Hα emission and colour observed over 80 kpc long tail of another target RB199 in Coma cluster, Yoshida et al. (2008) argues that the optically blue fireballs farther away contain a younger stellar population compared to fireballs near the parent galaxy. This requires further observational confirmation and careful interpretation as this is in contradiction to the natural expectation. The gas-clumps freshly stripped should have formed stars only recently and on the other hand clumps farther away, stripped earlier, should have relatively old stars as they have been forming stars for a longer time. It is interesting to note that such a 20-80 kpc long tail is like a "chart recorder" of star formation parameters, gradually changing along the length of the tail as the parent galaxy moves with a few thousand km s$^{-1}$ through the ICM. With the time since decoupling from the parent galaxy, the relative velocity of the stripped gas clouds and ambient medium should change from nearly a thousand km s$^{-1}$ to nearly zero, at the end of the tail where it joins the ICM, a point of no return. Possibly due to the detection of the H$\textsc{i}$ gas cloud near the end of the tail, Kenney et al. (2014) had postulated that once stars form in the striped gas cloud, those stars no longer feel the ram pressure and are left on the trail to age. On the other hand the gas cloud continues to be accelerated away from the parent galaxy due to ram pressure and continues to form stars. This led to the observation of stellar trails behind each fireball and younger stars farther away from the parent galaxy. However, such a simple-minded explanation is not satisfactory to us, as very good H I and Hα imaging of other ram pressure stripped galaxies show continuous stripping of gas clouds from the parent galaxy. Each fireball (gas clump) may show this age/colour gradient on kpc-scale but on a larger 10-100 kpc scale this should be washed out. The most detailed investigation for 80-kpc long tail and its fireballs exists for the Coma cluster galaxy RB199 (Yoshida et al. 2008). Contamination from unresolved background sources (also discussed by Kenney et al. 2014) and the inter-fireball light may contribute to the confusing trend of farther away fireballs being bluer. It is also known that the inter-fireball filamentary or diffuse emission is redder than the fireballs (Hester et al. 2010, Yoshida et al. 2008). Hence, a higher angular resolution imaging than achieved by GALEX is going to provide more accurate colour, star formation rate and age estimation of fireballs. We may expect the fireballs to remain compact, be resolved to sub-clumps, show head-tail structure or even show bow-shock at the head as seen in the case of Mira-A star with the GALEX (Martin et al. 2007). With the new result obtained from our high resolution ASTROSAT study, we propose an alternate hypothesis on the dynamical evolution of these Jellyfish galaxies with young star forming fireballs.

**2. Observation and Data Analysis:**
IC3418 was observed in the Far-UV (1300–1800 Å) using the Ultra-Violet Imaging Telescope (UVIT) onboard ASTROSAT (Agrawal 2006). The observations were carried out during May 2018. The galaxy was observed in two FUV filters F148W and F154W centred at 1481 Å and 1541 Å (Tandon et al. 2017), respectively. The log of observations are given in Table 1.

Table 1: UVIT Filters used in observation

| Filter | Mean Wavelength (Å) | Bandwidth (Å) | Exposure time (s) |
|---|---|---|---|
| F148W | 1481 | 500 | 2927.940 |
| F154W | 1541 | 380 | 1757.087 |

The images were processed at the UVIT-payload operation center (UVIT-POC) and made available to the investigators through the Indian Space Science Data Center (ISSDC). We carried out our analysis directly on the science-ready images made available to the ISSDC. The FUV images obtained are shown in Fig. 1. The images revealed five clumps of star formation or fireballs in the ram pressure-stripped tail of this Jellyfish galaxy. The AB magnitude of these clumps was measured using the latest photometric calibrations of UVIT given in Tandon et al. (2020). An aperture of 3" (~7 UVIT sub-pixels) centred at the local maxima were used for this purpose. This will contain more than 80% of the energy from the source (Tandon et al. 2020). The position and UV magnitudes of the clumps, in both the filters, are given in Table 2.

**Table 2: Photometry results of individual clumps**

| Clump | R.A. | DEC. | Mag. in F148W | Error | Mag. in F154W | Error |
|---|---|---|---|---|---|---|
| 1 | 12:29:47.56 | 11:23:35.67 | 22.035 | 0.283 | 21.812 | 0.384 |
| 2 | 12:29:49.66 | 11:23:08.58 | 21.812 | 0.256 | 21.778 | 0.378 |
| 3 | 12:29:50.73 | 11:22:35.01 | 21.553 | 0.227 | 21.458 | 0.326 |
| 4 | 12:29:52.73 | 11:22:49.21 | 21.368 | 0.208 | 21.321 | 0.306 |
| 5 | 12:29:52.03 | 11:22:04.74 | 21.644 | 0.237 | 21.444 | 0.324 |

## 3. Results

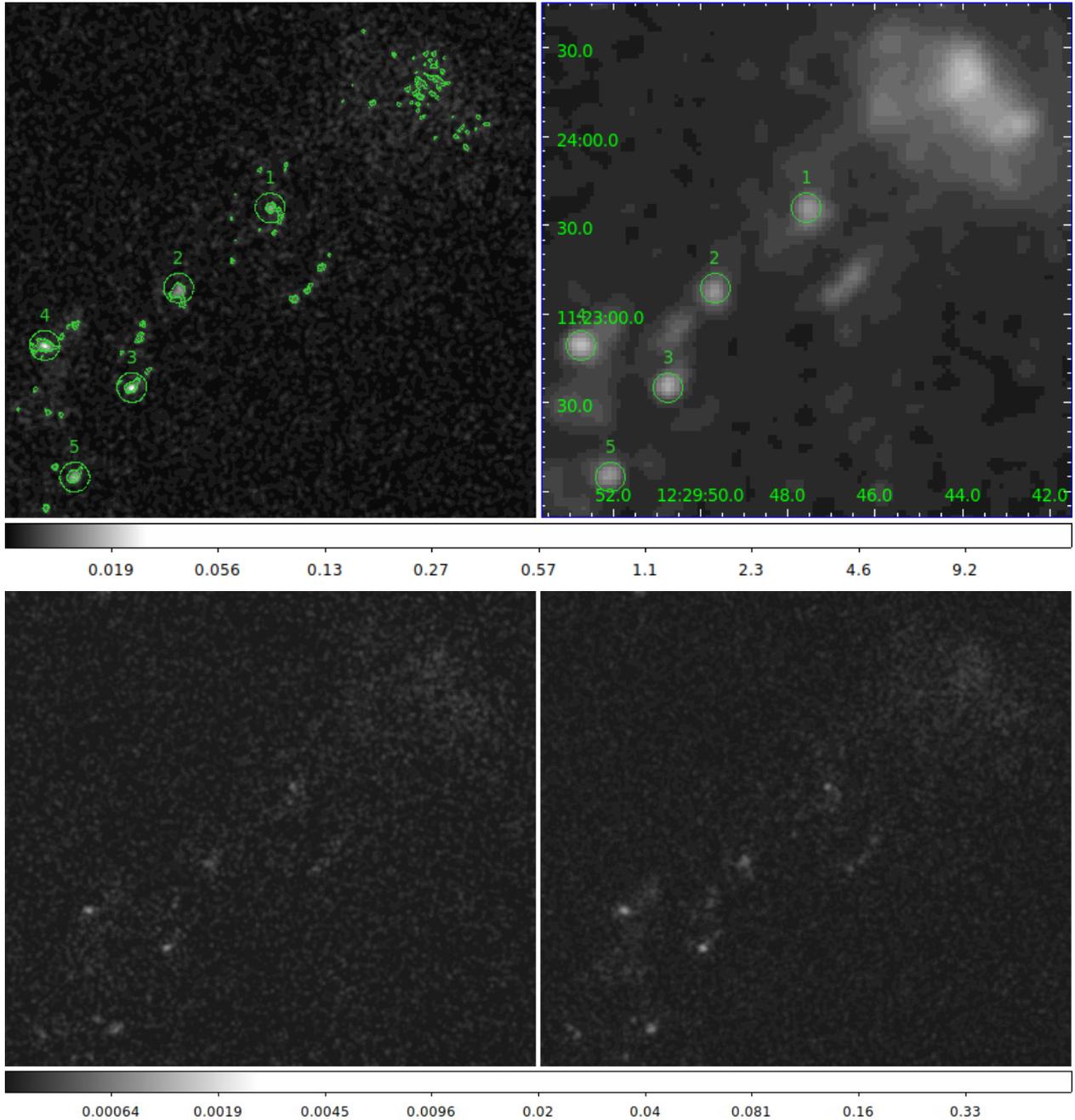

*Figure 1: shows both the GALEX NUV image (top right panel) and ASTROSAT UVIT far UV (F148W) image in both logarithmic grey scale and in green contours (top left panel). The images also label all the five compact star forming clumps (fireballs) that have been analysed in detail. The bottom panel shows the FUV image from ASTROSAT with F148W on the right and F154W on the left. Both the images are matched for scale, limit and colour bar for precise comparison.*

**IC3418:** Figure. 1 (top panel) shows the comparative high and low resolution structure of IC3418 as seen in ASTROSAT FUV (left) and GALEX NUV (right) images, respectively. In the bottom panel far-UV images from ASTROSAT, in both the filters, are presented. The FUV images, matched for scale, limit and colour bar, clearly show that higher significance of F148W image which has been used in all subsequent images. Here it is interesting to note that unlike the NUV GALEX image, the ASTROSAT FUV emission from the disk of the galaxy is getting fainter/resolved out at higher resolution compared to the compact clumps being analysed here. We have chosen to ignore the parent galaxy in our current study. Further, we have chosen not to present analysis on the diffuse kpc-scale star formation seen in this 17 kpc long tail which is detected clearly in the GALEX images along with these five compact clumps. However, note that the diffuse emission is redder in FUV-NUV colour compared to the compact clumps. This is clear from the photometric measurements presented in Table 1 by Hester et al. (2010). They find the average FUV-NUV colors for the diffuse emission is 0.35 ± 0.06 and for the fireballs to be -0.02 ± 0.03. The F148W-F154W colour as measured from ASTROSAT UVIT for fireballs 1 to 5 are 0.223, 0.034, 0.095, 0.047, 0.200 magnitudes, respectively (see Table 1 in this manuscript). Hence, though the wavelengths are very close, the colour gradient has not changed its sign in any case, suggesting that even at the highest angular resolution possible they are young star forming clumps and no other shock-related unusual process (Martin et al. 2007). We did not attempt photometric measurements of the diffuse emission but focus on the comparative analysis of the new structures that can be seen in the compact clumps (mentioned in the literature differently as knots or fireball) and have compared that with UV-optical colour images from HST archive.

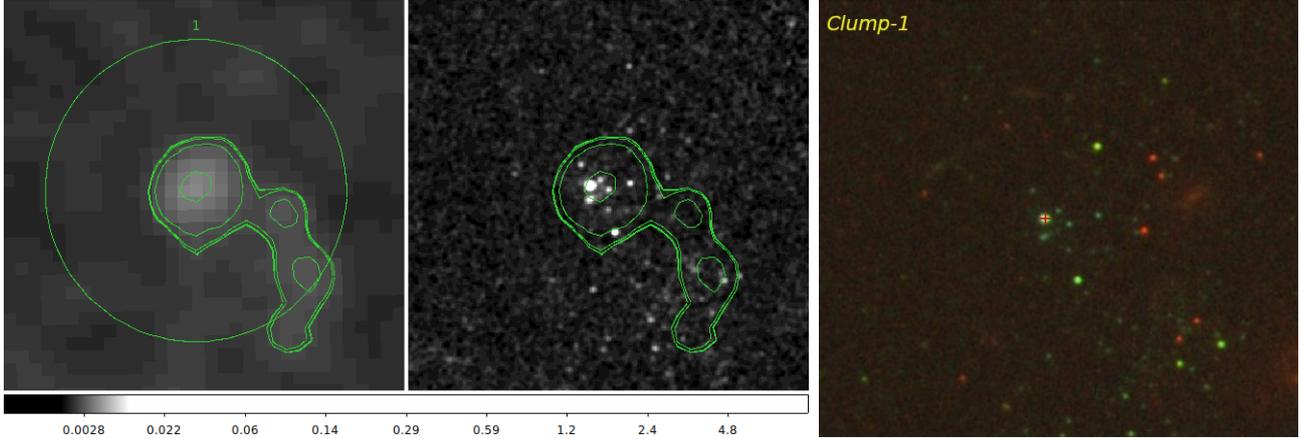

Fig. 2: Images presented here are for the Clump-1 (fireball-1), Left panel shows the UVIT FUV (F148W) image both in greyscale (logarithmic scale) and contours. The region-circle has a radius of 5'' and centered on the R.A. and Dec. for which photometry has been performed as described in Table 1. Middle panel presents the same contour map along with UV images from HST obtained from Hubble Legacy Archive (Proposal ID: 12756). The right panel represents a colour image of a slightly larger region around the centre of the clump. The red '+' mark in the colour image represents the centre of the left panel. The colour image has been created from the same HST data with blue, green and red showing F275W (effective wavelength = 2707.94 A), F475W (4746.16 A) and F814W (8044.99 A) filter images, respectively. The exposure times for the blue, green and red filter images are 2646, 2913 and 1356 seconds, respectively. Data presented in Fig. 2 to Fig 6 are identical except that they are for different clumps starting from Clumps-1 to Clump-5.

**Clump1:** Figure 2. Presents zoomed in view of the UVIT FUV images of Clump-1 (fireball-1), the closest to the parent galaxy, along with the UV image from HST. The contours from UVIT FUV (F148W) images are superposed on HST UV images for better comparison. The circles are centred on the compact clumps for which photometric measurements have been done on UVIT images. Note that the circles have a radius of 5 arcseconds which is about the angular resolution of GALEX images. The UVIT image showing compact central emission and diffuse emission seen on the south and south-west are consistent with the GALEX image. Note that our higher resolution images did not reveal any head-tail or bow-shock structure which were expected in certain models for how stars are formed in this exotic dynamical system. The bright compact clump in UVIT has multiple components as demonstrated in UV image from HST. Nearly 8 point sources can be noticed in the HST image which are located in the unresolved UVIT peak. The faint resolved point sources seen in the HST image located to the south and south-west of the core of the clump explain the faint UV extension seen in the UVIT image. Bottom panel with colour image from HST clearly shows further diffuse star formation. It is unclear if the red point sources belong to the tail of IC3418 or background galaxies. The extended red sources are definitely background galaxies. As the diffuse UV emission observed in GALEX is redder, some red point sources may be contributing to it. Thus clump-1 is compact in GALEX, and continues to be compact in UVIT but HST resolves that to a similar bright core consisting of nearly 8 surrounding point sources and scattered point sources to the south and south-west. Notably, the UVIT study did not show the head-tail or the shock-shell-like extended structure as expected in some models for fireballs. However, HST resolves them to point sources, likely to be single resolved stars.

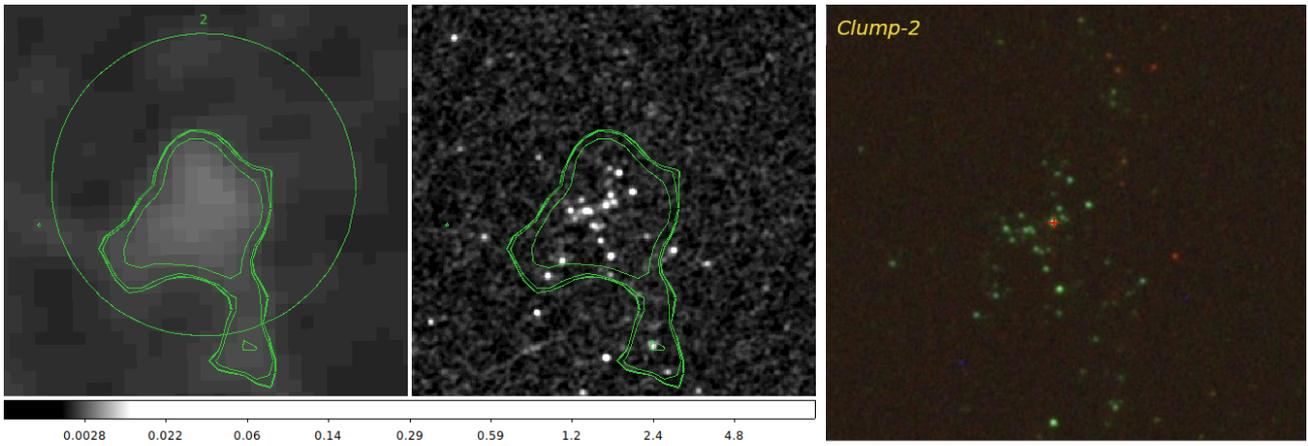

Figure 3. Imaging data presented here in all three panels are exactly the same as in Figure.2 but for the Clump-2.

**Clump-2:** In this compact clump UVIT seems to be resolving to multiple components with roughly equal brightness. HST resolves the structure further to a dozen of point sources that too conserve the overall structure. As this clump really lacks a prominent core, it is likely that all the point sources seen in HST images are single stars (similar to the claimed case of SDSS J122952.66 +112227.8, a blue supergiant star by Ohyama & Hota (2013) presented in later section). The colour HST image did not show any red point source in the compact clump but only farther away which may belong to the diffuse star formation which got possibly decoupled from the clump due to ram pressure (see fireball model figure 16 of Kenney et al. 2014). In summary, clump-2 (fireball-2) lacks a compact core, head-tail or bow-shock structure.

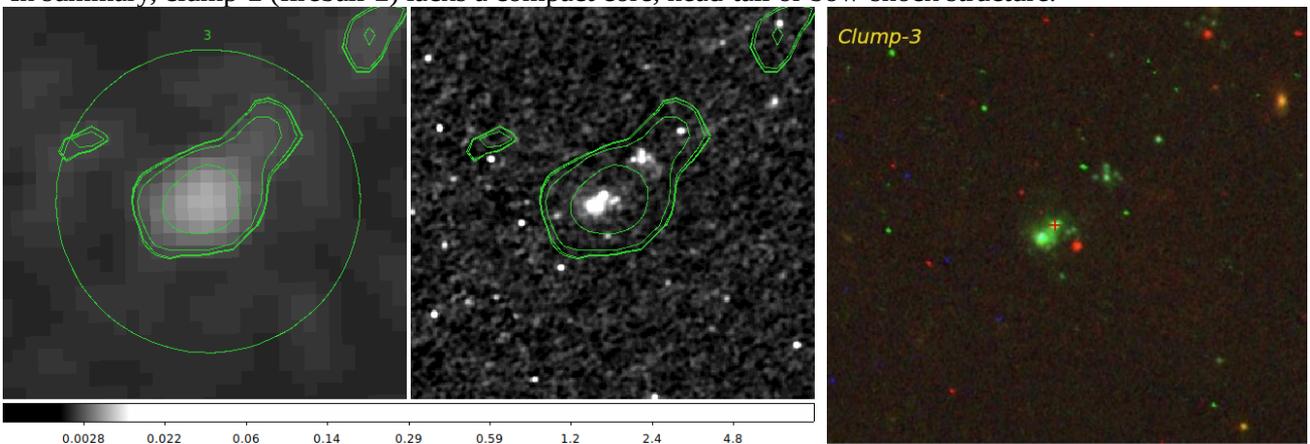

Figure-04: Imaging data presented here in all three panels are exactly the same as in Figure.2 but for the Clump-3 (fireball-3).

**Clump-3:** This compact clump in the GALEX image is resolved to a central peak and a tail to the north-west in the UVIT image. Here the tail extends towards the parent galaxy in the north-west. HST image finds bright point sources at the core with a sub-clump and a single point source making the tail. Also notice the diffuse emission seen in the clump. Within 5 arcsecond radius, a fainter but isolated point source could also be observed. Hence the compact core remains compact with close-enough point sources and a tail of sub-clump emission. Diffuse emission seen in the clump, in both UV and optical bands, may be scattering and/or unresolved stars.

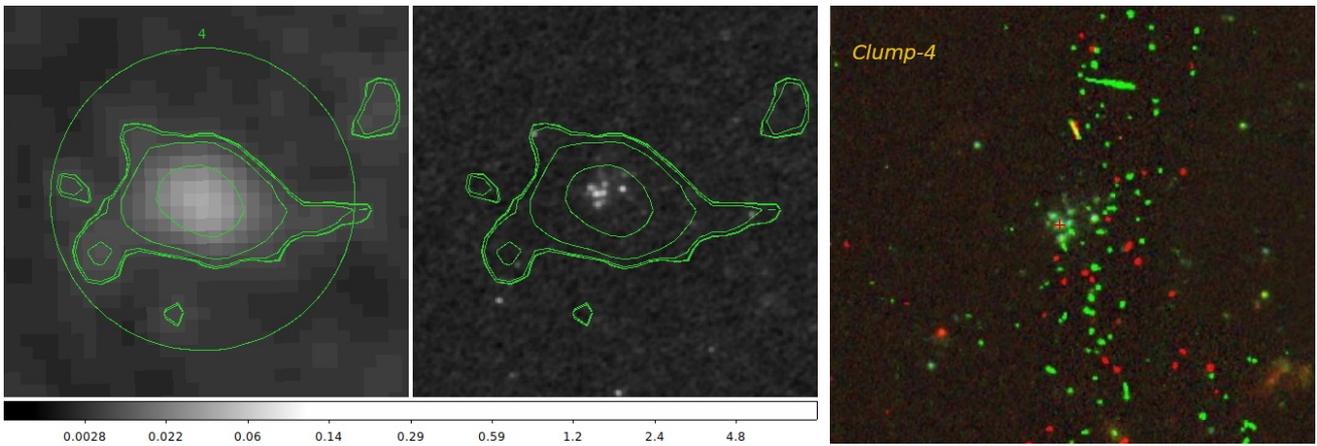

Figure. 5: Imaging data presented here in all three panels are exactly the same as in Figure.2 but for the Clump-4 (fireball-4). It is important to note here bad pixels running roughly north-south in a band. They appear in red and green channels only and HST UV image (blue) is not affected ( see middle panel).

**Clump-4:** This compact clump has been resolved by UVIT to be an elongated structure, roughly in the east-west direction. The HST UV image does not find any sub-clumps in the elongation but the core has closely-packed multiple point sources lacking any dominant member. The east-west elongation of the clump-4 can be resolved to point sources in the HST image.

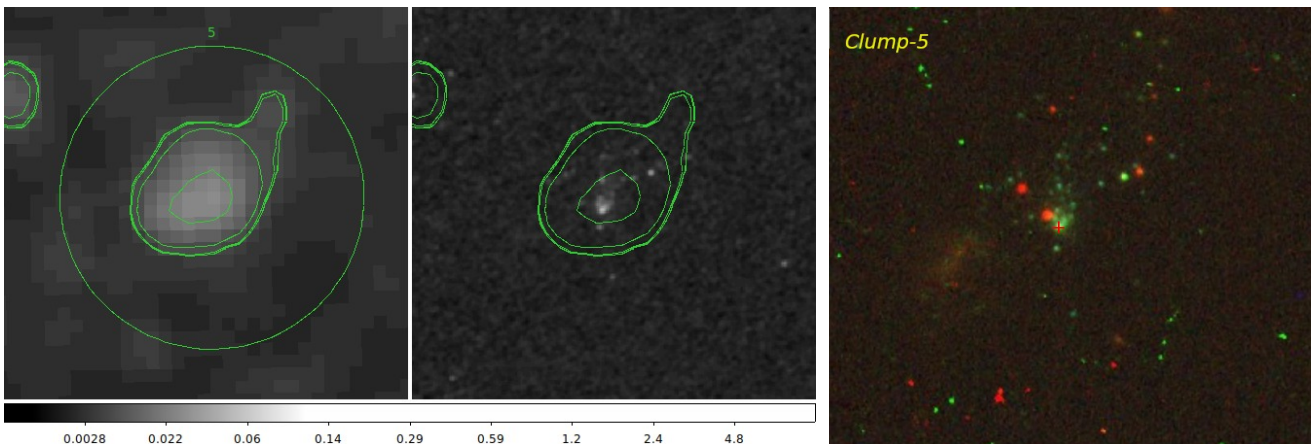

Figure 6: Imaging data presented here in all three panels are exactly the same as in Figure.2 but for the Clump-5 (fireball-5).

**Clump-5:** This is the farthest clump that we have analysed. Here again the central peak is prominent with possible extension in the direction of the parent galaxy. The HST UV images resolve the peak to be closely-spaced multiple point sources with a dominant member at the peak. The fainter point sources located on the north and west makes the UV tail observed in UVIT. Hence the compact core of GALEX remains a compact core in UVIT as well as HST images and the head-tail structure is also confirmed. Red optical point sources visible close to the peak and extended diffuse sources, either background or foreground, must be contaminating all measurements estimating colour and stellar populations.

**Blue Supergiant Star (BSG):** Finally, after these clumps/fireballs we focus our analyses with the star SDSS J122952.66+112227.8 which we had discovered in the diffuse star forming region between Clump-4 and Clum-5 and presented it to be a possibly single blue supergiant star (Ohyama & Hota 2013). The precise location of this BSG derived from UV images taken with the HST is R.A. 12:29:52.695 Dec. +11:22:27.937. We had noted a point source 1" away on the south-east and that is now resolved into two point sources with some diffuse emission. As presented in Figure 7 in the UV image (right panel) the BSG indeed is seen isolated without any sub-structure and strongly supports our claim of a single BSG star. Except the bottom one which is seen only in UV and thus likely an artifact, all the point sources have counterparts in both green and red optical filter images with the HST. Unfortunately due to the band of bad pixels HST source catalogue does not list its magnitudes. Comparing the pixel value (0.075) of the peak for BSG seen in UV image taken with the HST we notice that several of the point sources seen in each of these five compact clumps (fireballs) have similar or higher than this pixel value. Hence, we speculate that there would possibly be dozens of BSGs in this 17 kpc long tail of IC3418 which is at ~16.7 Mpc away. Since most of these compact fireball clumps are not having red point sources or diffuse/unresolved emission but resolved point sources they are likely well-separated O and B type stars. As ram pressure does not affect stars but only gas in the clumps, stars formed at different times may appear at different locations in the tail, acting like a chart-recorder of evolution of star formation. Some red point sources seen in HST images may indeed be such decoupled stars from a compact clump on the left. It is important to note a caveat in this picture that the radial velocity measured for this BSG is -99 km s$^{-1}$ which is different from radial velocities measured from the clumps in the tail. Though Ohyama & Hota (2013) have argued the physical association of the BSG with the tail of IC3418, based on luminosity and line ratio arguments, Kenney et al. (2014) doubt it based on these velocity measurements. With the HST imaging presented here where the BSG lacks any surrounding diffuse emission and many such point sources can be seen in the fireballs, the circumstantial evidence of the association seems strong compared to the radial velocity argument.

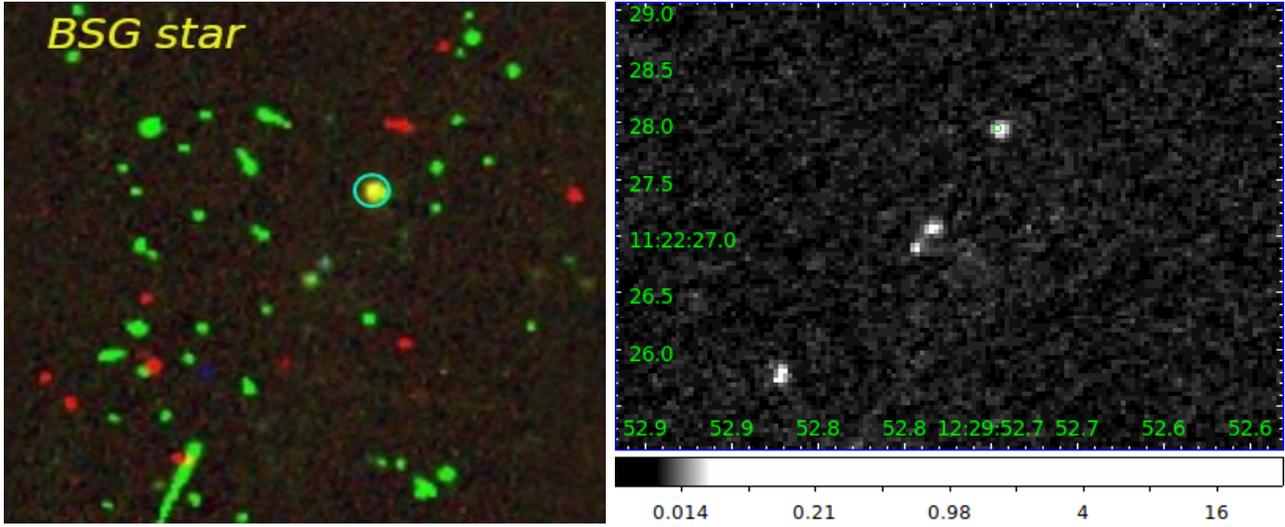

F*igure 7: Right panel shows the zoomed in UV image of the region around the blue supergiant (BSG) star SDSS J122952.66+112227.8 (topmost point source) from the UV HST data which is presented in grey logarithmic scale. The left panel is the same as in previous colour images from HST. As described in Figure 5 bad pixels affect the HST colour map. The BSG has been labelled with a cyan circle. The blue point source (left panel), also seen as the bottommost point source in the (right panel), is likely an image artifact which has no counterpart in green and red filter images. Rest all three point sources in the UV image have optical counterparts in HST data.*

**4. Discussion:** Similar to the five compact clumps or fireballs that we have defined here, nearly six regions of diffuse star formation can be seen. Hester et al. (2010) defined 9 clumps (knot K1 to K9) and 3 diffuse regions (D1, D2, D3). Note that diffuse regions are in general redder and possibly contain old stars likely because they are already decoupled from star forming gas clouds due to difference in ram pressure that a molecular gas cloud and a compact star would experience. Such red colour in optical is also seen in RB199, the next nearest Jellyfish galaxy in the Coma cluster (Yoshida et al. 2008). Although the tail is expected to have a lower surface density of gas than the galaxy disk itself, the higher resolution of ASTROSAT shows the fireballs to be brighter in FUV than the disk (Fig. 1). It is unclear if higher extinction in the parent galaxy is the reason. However, neutral H I gas cloud is detected at the end of the tail and not from the parent galaxy (Kenney et al. 2014). A suitable mechanism of creating high density molecular gas clouds must be existing in these compact clumps to emit bright UV emission from young star formation. On the contrary, ionisation/heating and dispersion of the stripped diffuse cold neutral gas cloud would also be unavoidable coming in contact with the hot ICM plasma in high relative velocity.

**Fireball Evolutionary Sequence:** We looked for possible gradual change in the measured physical parametres along the tail, with the distance from the parent galaxy. The star formation rate (SFR) is estimated following the prescription for nearby galaxies as in the Karachentsev et al. 2013. The FUV (F148W) magnitudes from the Clumps/fireballs, as listed in the Table-2, has been used. Galactic extinction coefficient (A=0.248 mag.) was applied following Cardelli et al. 1989 and no correction for internal extinction was assumed as fireballs are far away from the dwarf parent galaxy, which has already been stripped of its HI

ISM. We find that the SFR in the clumps/fireballs ranges from 4 to 7.4 x $10^{-4}$ M$_\odot$ yr$^{-1}$ with the error ranging 1-4 x $10^{-4}$ M$_\odot$ yr$^{-1}$ (see Table 3). This is comparable to the values found by Fumagalli et al. (2011) estimated using Hα flux densities. This is an order of magnitude less than that for fireballs in the 80 kpc long tail of RB199 (Yoshida et al. 2008). It can be noticed clearly that the FUV SFR of fireballs increases with the distance from the parent galaxy (Table 3 and Figure 8). From figure we can see an increase of ~1 x $10^{-4}$ M$_\odot$ yr$^{-1}$ of SFR per arcmin (4.8 kpc) projected distance. We further expressed it in optical image with other measured parametres of the fireballs (Figure. 9). This increase in SFR can explain the observation of fireballs farther away showing relatively bluer optical and UV colour in various observations (Hester et al. 2010, Yoshida et al. 2008, Fumagalli et al. 2011, Kenney et al 2014). As discussed earlier, the actual reason behind this is younger age of stars at the tail end or is due to higher extinction close to the parent galaxy or the difference in metallicity of stripped gas spread on to the tail could not be disentangled. However, at least in the case of IC3418, H I gas cloud is found near the tail end and Hα emission is seen at the tail end, ruling out extinction as a major reason. Furthermore, IC3418 is a dwarf galaxy and a proper nucleus could not be figured out, which would disfavour the difference in metallicity. No gradient of metallicity was observed in the tail (Kenney et al. 2014). Hence, higher SFR of fireballs farther down the stream naturally explains the bluer colour of fireballs farther from the galaxy. This was not identified by previous studies possibly because they could not focus on the compact fireballs.

We have further correlated the observed radial velocities of fireballs available in the Kenney et al. (2014). Direct measurements are not available for fireball-1 and fireball-2, hence, the velocity of the nearest region has been assumed. The radial velocity of the parent galaxy (176 km s$^{-1}$) has been subtracted from the fireball velocity. Higher the fireball-galaxy velocity difference, lower would be the fireball-ICM velocity difference. Hence, as expected, the fireball-4 with the highest SFR has the highest fireball-galaxy velocity difference (117 km s$^{-1}$) or least fireball-ICM velocity difference. We further have collected the stellar population age of the fireballs from Fumagalli et al. (2011). Any obvious trend is not seen here. If fireballs continue to form stars and they indeed get decoupled after forming, a large-scale age gradient is likely to be not observed that easily. Hence SFR gradient seems to be the only explanation of the UV-optical colour gradient observed.

**Table 3. Properties of the fireball clumps**

| Clump No. | Distance from galaxy in arcmin | FUV SFR ($10^{-4}$ M$_\odot$ yr$^{-1}$) | Error SFR ($10^{-4}$ M$_\odot$ yr$^{-1}$) | Fireball-galaxy velocity difference in km s$^{-1}$ | Stellar Age in M yr |
|---|---|---|---|---|---|
| Fireball-1 | 1.13 | 4.008 | 1.045 | 60 | 130 |
| Fireball-2 | 1.83 | 4.921 | 1.160 | 42 | 80 |
| Fireball-3 | 2.39 | 6.247 | 1.306 | 44 | 390 |
| Fireball-4 | 2.55 | 7.408 | 1.419 | 117 | 170 |
| Fireball-5 | 3.00 | 5.745 | 1.254 | 49 | 740 |

**Vortex street model:** Various simulations of ram pressure stripping of cluster infalling galaxies have shown turbulent gas in the wake of the parent galaxy (Kapferer et al. (2009), Tonnesen & Bryan (2012), Steinhauser, Schindler & Springel, (2016), Muller et al. (2020)). The cold neutral atomic/molecular gas that is stripped from the galaxy experiences the million degree hot plasma of the ICM with significantly less density than the stripped gas. The Kelvin-Helmholtz instabilities that would be developing due to the velocity shear in between these two fluids are going to create vortices. Under self-gravity these vortices should be forming dense gas clouds collapsing to form stars. Furthermore, on large-scale, gas streams/clouds closer to the galaxy would have higher relative velocity with the ICM in the immediate vicinity of the parent galaxy than gas located farther down the stream of the tail which ideally should diminish to zero difference and eventually decouple from the galaxy. On large 10 to 100 kpc scales, such structures can be compared with the von Karman vortex street that forms in hydrodynamic flows around an obstacle (in relative terms, ICM is flowing past the galaxy with ISM as the dye injected to visualise the vortex). Such vortices grow with distance from the obstacle and gradually disperses in the ambient medium (Etling 1989, Horváth et al. 2020). Unlike terrestrial fluid flow experiments, here the vortices formed would be trying to nucleate or move radially inward under self-gravity and turbulence trying to compete with and disperse it down the stream. Hence with the decrease of relative velocity downstream, gravity can take over turbulence and the vortices can continue to collapse gas clouds and form young stars. This competing process can be seen on a larger 10-100 kpc scale along the tail and can explain the puzzling trend in Jellyfish galaxies of why Hα emission and optically-blue or UV-bright star formation is seen in clumps/fireballs downstream than close to the parent galaxy. Although molecular gas (CO emission line) has been detected, thanks to the amazing sensitivity of ALMA, from yet another stunning Jellyfish galaxy (Jachym et al 2019), it is yet to image a full velocity field to confirm/inform dynamical model of vortices for fireballs that we hypothesise here. Cold molecular gas observations have detected the galaxy but not from the tail (Jáchym et al. 2013). Here we propose that fireballs are the vortices where with the passage of time (with distance from parent galaxy), the self-gravity takes over turbulent forces (proportional to fireball-ICM velocity difference) and cold gas-density grows leading to higher SFR (expressed as blue fireballs). As seen in Figure 8, SFR may increase to a certain distance and then decline. Further, recent magnetic field observations with the Karl G. Jansky Very Large Array (VLA) have revealed highly ordered magnetic fields in the tails of several Jellyfish galaxies and resolution to investigate vortices in fireballs/clumps are yet to be achieved (Muller et al. 2020). IC3418 has very faint radio continuum detection and reasonable radio continuum imaging with polarisation information may require the next generation of radio telescopes like the Square Kilometre Array (SKA).

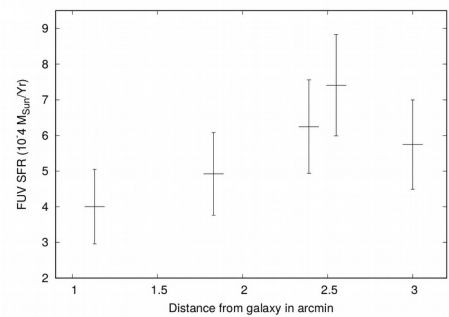

*Figure 8: Plot of the star formation rate, along with the error, has been plotted with distance of the fireballs from the parent galaxy.*

**5. Conclusion:**
1. With ASTROSAT UVIT far-UV image resolution higher than the previously available GALEX UV images of this nearest Jellyfish galaxy we have resolved the compact star forming clumps (fireballs) to sub-clumps.
2. Incorporating UV and optical images available in HST archive, we resolve these fireballs to several point sources seen in UV.
3. With the HST images, we confirm the isolated nature of the previously discovered farthest star, a blue supergiant star ( SDSS J122952.66+112227.8) located in the same tail, and speculate the presence of a dozen such individual stars in the fireballs.
4. We find that the FUV SFR of fireballs increases with the distance from the parent galaxy.
5. We propose a model where the fireballs are the vortices that develop due to Kelvin-Helmholtz instability due to velocity shear between cold-dense stripped gas and hot rare gas/plasma of the intra-cluster medium.
6. The gas clouds in the far away fireballs are actually old compared to that in the fireballs near the galaxy, as they were stripped from the galaxy earlier. Thus, fireballs farther away showing blue optical/UV colour or possibly younger stellar population is counter-intuitive. This can be easily explained in our vortex model where self-gravity takes over turbulence as relative velocity decreases with time or distance from the parent galaxy along the turbulent wake, resulting in higher SFR in the fireballs at the far end of the tail.
7. Such targets can be unique laboratories to understand basics of star formation as well as various high-energy phenomena due to the presence of massive stars well-separated from each other and arranged in chronological order due to ram pressure stripping.

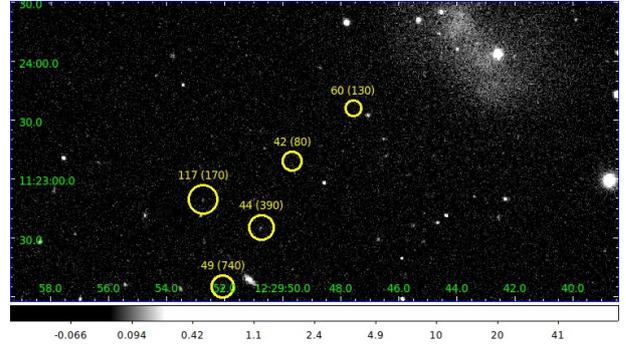

*Figure 9: The SDSS r-band image shows yellow circles around the fireballs with size proportional to FUV SFR and are labelled with velocity difference ( age of stellar population ). See Table 3 and Figure 8 for more data.*

**Acknowledgement:** AH is thankful to the University Grants Commission (India) for the one-time start-up and monthly salary grants, under the Faculty Recharge Programme. This publication uses the data from the ASTROSAT mission of the Indian Space Research Organisation (ISRO), archived at the Indian Space

===